\documentclass[a4paper,aps,prl,showpacs,twocolumn,floatfix,superscriptaddress]{revtex4}
\usepackage{graphicx} 
\usepackage{amsmath}
\usepackage{bm}
\usepackage{dcolumn}

\begin{document}

\title{Magic Wavelength for Atomic Motion Insensitive Optical Lattice Clocks}

\author{Hidetoshi Katori}
\affiliation{Department of Applied Physics, Graduate School of Engineering, The University of Tokyo,
Bunkyo-ku, Tokyo 113-8656, Japan}
\affiliation{CREST, Japan Science and Technology Agency, 4-1-8 Honcho Kawaguchi, Saitama 332-0012, Japan}

\author{Koji Hashiguchi}
\affiliation{Department of Applied Physics, Graduate School of Engineering, The University of Tokyo,
Bunkyo-ku, Tokyo 113-8656, Japan}
\affiliation{CREST, Japan Science and Technology Agency, 4-1-8 Honcho Kawaguchi, Saitama 332-0012, Japan}
\author{E.~Yu.~Il'inova}
 \affiliation{Physics Department, Voronezh State University, Universitetskaya pl.1, Voronezh 394006, Russia}

 \author{V.~D.~Ovsiannikov}
 \affiliation{Physics Department, Voronezh State University, Universitetskaya pl.1, Voronezh 394006, Russia}

\date{\today}
\begin{abstract}
In a standing wave of light, a difference in spatial distributions of
multipolar atom-field interactions may  alter the definition of the magic wavelength to minimize the uncertainty of  optical lattice clocks.
We show
that the magic wavelength can be determined so as to eliminate the
spatial mismatch in electric dipole, magnetic dipole, and electric quadrupole interactions for specific combinations of standing waves.
Experimental prospects of such lattices used with a blue magic
wavelength are discussed.
\end{abstract}

\maketitle
Quantum absorbers trapped in  well-designed electromagnetic fields are the excellent candidates for future optical atomic clocks projecting uncertainties exceeding $10^{-18}$~\cite{Deh82,Kat03}, which represent the state of the art of the precision spectroscopy~\cite{Ros08,Lud08}.
While these atom traps provide long interrogation time and the Lamb-Dicke confinement of atoms necessary for  ultrahigh resolution  spectroscopy,  the relevant trapping fields  impose an inherent fundamental limit on  measurement uncertainties due to the atomic multipolar interactions with trapping fields~\cite{Deh82,Yu94,Ita00,Tai06} and  hyperpolarizability effects~\cite{Kat03,Bru06}. 
It is of note that the detection and control~\cite{Bar04,Dub05,Roo06} of the electric quadrupole interactions of ions with their trapping fields triggered an essential breakthrough for ion clocks operating on the $S-D$ clock transitions to achieve the the uncertainties of   
 $10^{-15}$ and even below~\cite{Sch05,Ros08}.  

Recently it was pointed out that the multipolar interactions of atoms with optical lattices may introduce spatial mismatch of the lattice potentials in the clock transition thus affecting optical lattice clocks' uncertainties at  $10^{-16}$~\cite{Tai08}.
This inferred the slight breakdown of the original concept of the magic wavelength that cancels out the quadratic light shift in the clock transitions~\cite{Kat03}.
In this Letter, we discuss strategies to minimize the light shift uncertainties in optical lattice clocks
by considering the electric dipole (E1), magnetic dipole (M1), and electric quadrupole (E2) interactions of atoms in a standing wave.
For specific lattice geometries, we show that the magic frequency $\omega_{\rm m}$ of
a lattice clock can be defined so as to eliminate spatial mismatch of the lattice potentials, which is accompanied by a spatially constant differential light shift $\delta \nu$.
Consequently, optical lattice clocks free from atomic-motion-dependent clock shift are
realized and the offset $\delta \nu$ can be evaluated down to $10^{-18}$ as a correction for the clock frequency.
We discuss experimental prospects for the  $^1S_0-{}^3P_0$  clock transition of Sr atoms.
In particular, combined with a blue-detuned magic wavelength
\cite{Tak09},  the proposed lattice geometry closely
simulates the Paul trap employed in  ion clocks \cite{Deh82}, thus pushing lattice clocks' uncertainty towards the $10^{-18}$ regime.

\begin{figure}[b]
\begin{center}
\includegraphics[width=0.9\linewidth]{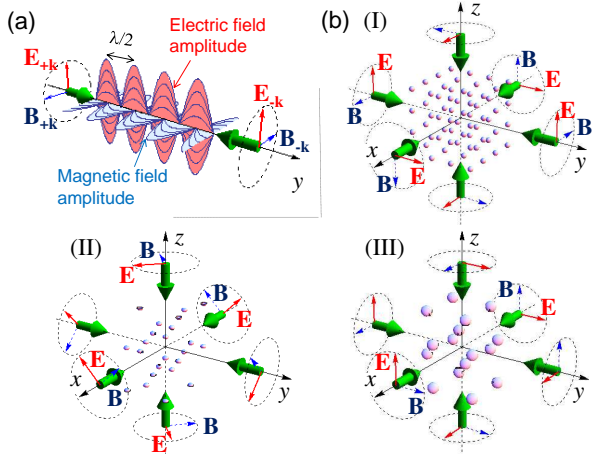}
\caption{(a) Spatial distribution of an electromagnetic field for a 1D standing wave. (b) Configuration of the electromagnetic fields and optical lattices for the cases (I)-(III), as described in the text. Optical lattice sites inside  $|x|,|y|,|z|<0.9\lambda$ are indicated with their  equipotential surfaces  given by $q_{\rm E1}({\bf{r}})=0.3$ and $\rho_\xi=1$.}
\label{FIG:lattice}
\end{center}
\end{figure}

Optical lattices consist of a spatially periodic light shift
formed by an interference pattern of  electromagnetic waves.
The simplest configuration is a one-dimensional (1D) lattice with a standing wave of
light having wave vectors $\pm {\bf k}$  and parallel electric
fields  ${\bf E}_{\pm \bf k}$ as shown in Fig.~\ref{FIG:lattice}(a).
The corresponding magnetic fields ${\bf B}_{\pm \bf k}$ are
proportional to $({\pm \bf k})$ $\times$ ${\bf E}_{\pm \bf k}$. As a
result,  anti-nodes of the total electric field ${\bf E}={\bf E}_{+ \bf k}+{\bf E}_{- \bf k}=2{\bf E}_{\pm \bf k}$
correspond to  nodes of the magnetic field ${\bf B}={\bf B}_{+ \bf k}+{\bf B}_{- \bf k}=0$, and their
amplitudes are a quarter of the wavelength $\lambda$=$2\pi/|{\bf
k}|$ out of phase, introducing different spatial dependences  for the E1 and M1 interactions.
Moreover, as
the E2 interaction is proportional to the
electric field gradient, its spatial dependence also differs from
that of E1.
These different spatial dependences do not allow perfect cancellation of the quadratic light shift in the clock transitions~\cite{Tai08}.
However, by admitting a constant differential light shift offset, we will show that spatial mismatch of the light shift can be eliminated, therefore the atomic motion dependent clock shift, which is detrimental to  atomic clocks, can be removed.

As there are hyperpolarizability effects~\cite{Kat03,Bru06} that
cannot be moderated by the magic wavelength,  a blue detuned
magic wavelength that confines atoms near the nodes of  standing
waves would be a promising choice, which will  reduce the relevant
uncertainties down to $2\times10^{-19}$~\cite{Tak09}. Although the
local electric field intensity for atoms is reduced, the E2
interaction, in turn, may severely affect the accuracy of a blue detuned lattice
clock, as the electric field gradient can generally be maximum at
the nodes of the standing wave.
These blue detuned lattices can be realized using  3D lattice configuration~\cite{Mue97}.
Moreover, application of  single-occupancy 3D lattices with red and blue magic wavelengths may effectively suppress the collision shift~\cite{Aka08}.

We consider 3D optical lattices consisting of three mutually orthogonal standing
waves, whose results are straightforwardly applicable to lower dimensional lattices.
In particular, we derive spatial dependences $q_X({\bf r})$ of the quadratic light shift with $X$~=~E1, M1, and E2 interactions.
For generality, we assume different electric field
amplitudes $E_\xi=\rho_\xi E_0$ of the three standing waves with
equal wavelength $\lambda$ and wave number $k$=$2\pi/\lambda$ and
different polarization vectors ${\mathbf p}_\xi$ and ${\mathbf
p}^b_\xi$ ($|{\mathbf p}_\xi|$=$|{\mathbf p}^b_\xi|$=1) for the
forward  and backward running waves, respectively, in the $\xi$=$x,y,z$ directions denoted by unit vectors ${\bf e}_\xi$:
\begin{equation}
\begin{split}
{\bf{E}}_\xi  (\xi ,t)
& = E_\xi  \left[ {{\bf{p}}_\xi  \cos (k\xi  - \omega t) + {\bf{p}}_\xi ^b \cos (k\xi  + \omega t)} \right]\\
&= E_\xi  \left( {{\bf{p}}_\xi ^ +  \cos k\xi \cos \omega t + {\bf{p}}_\xi ^ -  \sin k\xi \sin \omega t} \right),
\end{split}
\label{Efield}
\end{equation}
where we define ${\bf{p}}_\xi ^ \pm $$  \equiv $$ {\bf{p}}_\xi \pm
{\bf{p}}_\xi ^b$. The total electric field vector in this lattice
is given by ${\bf{E}}({\bf{r}},t)$=$\sum_{\xi  = x,y,z}
{{\bf{E}}_\xi  (\xi ,t)} $.

The
principal contribution to the lattice potential is given by the
second-order quasienergy shift due to the electric dipole
atom-field interaction determined by the Hamiltonian $\hat
V_{\rm E1}({\bf{r}},{\bf{r}}_e,t)$=$-$$\mathbf d\cdot\mathbf
E({\bf{r}},t)$ \cite{MOR86},
\begin{eqnarray}
\label{StE1}
U_{\rm E1}(\mathbf r)&=&-\langle\langle \psi|\hat V_{\rm E1}\mathcal
G(\mathbf r_e,t;\mathbf r'_e,t')\hat V_{\rm E1}|\psi\rangle\rangle
\nonumber
\\ &=&-\frac{E_0^2}{2}\alpha_{\rm E1}(\omega)q_{\rm E1}(\mathbf r),
\end{eqnarray}
where $\mathbf r_e$ is the position vector of the outermost atomic
electron relative to the  atomic nucleus at $\mathbf r$, $\mathbf
d$=$-\mathbf r_e$ is the electric dipole moment, and $\mathcal
G(\mathbf r_e,t;\mathbf r'_e,t')$ is the quasienergy Green
function of an  atom. Here the atomic units are used,
$e$=$m$=$\hbar$=1, where the speed of light is
$c$$\approx$137. The double angular brackets in
Eq.~(\ref{StE1}) denote the time integration (in variables $t$ and
$t'$) over the field oscillation period $T$=$2\pi/\omega$ and the
spatial integration over the position $\mathbf r_e$  of the atomic
electron.  With the
use of the electric field in Eq.~(\ref{Efield}), the spatial
distribution function of the atom-field E1 interaction energy is
given by
\begin{equation}
q_{\rm E1} ({\bf{r}}) = \frac{1}{2}\Bigl( {\sum\limits_\xi {\rho_\xi
{\bf{p}}_\xi ^ +  \cos k\xi } } \Bigr)^2  + \frac{1}{2} \Bigl(
{\sum\limits_\xi  {\rho_\xi  {\bf{p}}_\xi ^ -  \sin k\xi } }
\Bigr)^2. \label{E1}
\end{equation}

To evaluate the contribution of the M1 interaction, it is
sufficient to determine the magnetic field component of the
lattice, which for each running wave with  wave vector $\mathbf k_\xi$
is given by
${\bf{B}}_\xi({\bf{r}},t)$=$\mathbf k_\xi/k$$\times$$ {\bf{E}}_\xi({\bf{r}},t)$.
The total magnetic field corresponding to the electric field in Eq.~(\ref{Efield}) is given by,
\begin{eqnarray}
\mathbf B  ({\bf r} ,t)
 = \sum_{\xi=x,y,z} &E_\xi & \bigl(\mathbf e_\xi\times\mathbf p_\xi^+ \sin k\xi \sin \omega t \nonumber \\
& +&\mathbf e_\xi\times \mathbf p_\xi ^- \cos k\xi\cos \omega t
\bigr).
\label{Bfield}
\end{eqnarray}
The magnetic dipole contribution to the lattice potential may be
written similarly to Eq.~(\ref{StE1}), as the quasienergy shift
corresponding to the atom-field M1 interaction is described by the
Hamiltonian $\hat V_{\rm M1}$=$-$$\hat{\mathbf m}\cdot\mathbf
B({\bf{r}},t)$, where  $\hat{\mathbf m}$=$-(\hat{\mathbf
J}$+$\hat{\mathbf S})/2c$ is the magnetic moment of an atom with
 atomic total momentum ${\mathbf J}$ and spin ${\mathbf S}$.
The spatial distribution of the M1 interaction is given by,
\begin{equation}
\begin{split}
q_{\rm M1} ({\bf{r}})= & \frac{1}{2}\Bigl(\sum\limits_\xi  \rho_\xi  {\bf{e}}_\xi   \times {\bf{p}}_\xi ^+  \sin k\xi  \Bigr)^2  \\
&+ \frac{1}{2}\Bigl(\sum\limits_\xi  \rho_\xi  {\bf{e}}_\xi \times
{\bf{p}}_\xi ^ -  \cos k\xi  \Bigr)^2.
\end{split}
\label{M1}
\end{equation}

In the nonrelativistic approximation, the magnetic dipole
polarizability in the $n\, ^1\!S_0$ ground state  is zero, 
while for the $n\, ^3\!P_0$ metastable state,  it is given by
\begin{equation}\label{aM1}
\alpha_{\rm M1}(\omega)\!=\frac{E_{n\, ^3\!P_1}-E_{n\,
^3\!P_0}}{6c^2[(E_{n\, ^3\!P_1}-E_{n\, ^3\!P_0})^2-\omega^2]},
\end{equation}
which is evidently the value of the second order in the
fine-structure constant $\alpha$=$1/c$. $\alpha_{\rm M1}(\omega)$ remains negative for
 $\omega$ higher than the $n\, ^3\!P_0- n\, ^3\!P_1$  transition frequency.

Not less important than the M1 Stark shift may be
the contribution of the electric
quadrupole (E2) interaction,
\begin{eqnarray} \label{StE2}
U_{\rm E2}(\mathbf r)&=&-\langle\langle \psi|\hat V_{\rm E2}\mathcal
G(\mathbf r_e,t;\mathbf r'_e,t')\hat V_{\rm E2}|\psi\rangle\rangle
\nonumber
\\ &=&-\frac{E_0^2}{2}\alpha_{\rm E2}(\omega)q_{\rm E2}(\mathbf r).
\end{eqnarray}
The value of the quadrupole polarizability $\alpha_{\rm E2}(\omega)$
is of the second order in the fine-structure constant $\alpha$,
just as  the magnetic dipole polarizability in Eq.~(\ref{aM1}) is.
The E2 interaction operator  may be taken from the Taylor series
in powers of the small parameter $k|\mathbf r_e|$$\ll$1 for the
total atom-electric-field interaction Hamiltonian,
\begin{equation} \label{VEk}
\hat V_{\rm E}(\mathbf r,\mathbf r_e, t)=\mathbf
r_e\cdot\sum\limits_{s=0}^\infty\frac{(\mathbf
r_e\cdot\nabla)^s}{(s+1)!}\, \mathbf E(\mathbf r, t),
\end{equation}
where all derivatives are taken with respect to the components of
the position vector $\mathbf r$, while $\mathbf r_e$ is assumed to be
constant.
$\hat V_{\rm E}(\mathbf r,\mathbf r_e, t)$ includes all higher-order multipole interactions: The $s$=0 term    corresponds to the
Hamiltonian $\hat V_{\rm E1}({\bf{r}},t)$ and the  $s$=1 term to the Hamiltonian $\hat V_{\rm E2}({\bf{r}},t)$.
After substitution of this operator into Eq.~(\ref{StE2}) and
 integration over time and angular variables, the spatial
 distribution of the quadrupole energy  is
 determined:
\begin{equation}
\begin{split}
q_{\rm E2} &({\bf{r}}) = \frac{1}{2}\sum\limits_{(\xi ,\eta) }
\left( {\rho_\xi  {\bf{e}}_\eta   \cdot {\bf{p}}_\xi ^ +  \sin k\xi
+ \rho_\eta  {\bf{e}}_\xi   \cdot {\bf{p}}_\eta ^ +  \sin k\eta } \right)^2   \\
&+ \frac{1}{2}\sum\limits_{(\xi ,\eta) } \left( {\rho_\xi
{\bf{e}}_\eta \cdot {\bf{p}}_\xi ^ -  \cos k\xi  + \rho_\eta
{\bf{e}}_\xi   \cdot {\bf{p}}_\eta ^ -  \cos k\eta } \right)^2,
\end{split}
\label{E2}
\end{equation}
where the sum runs over $(\xi ,\eta)  = (x,y),\,(y,z)$, and
$(z,x)$. Correspondingly, the quadrupole polarizabilities of the
ground and excited states are written in terms of the radial
matrix elements, e.g., for  the $|\psi\rangle$=$|n\,^1\!S_0\rangle$  state,
\begin{equation}
\alpha_{\rm E2}^{^1\!S_0}(\omega)=\frac{\omega^2}{60c^2}\langle n\, ^1\!S_0|
 r_e^2(g^\omega_{^1\!D_2}\!+\!g^{-\omega}_{^1\!D_2})
 r_e^2|n\, ^1\!S_0\rangle,
\label{aE2g}
\end{equation}
where the radial Green functions $g_{^1\!D_2}^{\pm\omega}$ of the
singlet $D$-state subspace appear.

Below we illustrate a few representative examples that allow cancellation of spatial mismatch of the lattice potentials.
They are assorted by the  electric field  ${\bf{E}}_f$ and ${\bf{E}}_b $  of the forward and backward  running waves that compose lattice standing waves, as summarized in Fig.~\ref{FIG:lattice}(b).

(I) ${\bf{E}}_f ||{\bf{E}}_b $ standing waves (${\bf{p}}_\xi$=${\bf{p}}_\xi^b $), in which we take ${\bf{p}}_x$=$
{\bf{e}}_y$, ${\bf{p}}_y$=${\bf{e}}_z$, and ${\bf{p}}_z$=$
{\bf{e}}_x $. The E1 distribution is calculated to be,
\begin{equation}
q_{\rm E1} ({\bf{r}}) = 2\left( {\rho_x^2 \cos ^2 kx + \rho_y^2 \cos ^2
ky + \rho_z^2 \cos ^2 kz} \right). \label{caseI}
\end{equation}
The M1 and E2 distributions are given by,
\begin{equation}
q_{\rm M1} ({\bf{r}}) = q_{\rm E2} ({\bf{r}}) = \Delta q -
q_{\rm E1} ({\bf{r}}), \label{caseIb}
\end{equation}
with $\Delta q=2(\rho_x^2$+$\rho_y^2$+$\rho_z^2)$. Thus the distributions
of M1 and E2 shifts in this lattice coincide and differ from
$q_{\rm E1} ({\bf{r}})$ in sign and by a constant offset of
$\Delta q$.

(II) ${\bf{E}}_f  \bot {\bf{E}}_b $ standing waves (${\bf{p}}_\xi  \bot {\bf{p}}_\xi^b $) with their polarization vectors
pointing at an angle $\pi /4$ to the standing-wave beams, i.e.,
${\bf{p}}_x$=$({\bf{e}}_y $+${\bf{e}}_z)/\sqrt2$,
${\bf{p}}_x^b$=$(-{\bf{e}}_y $+${\bf{e}}_z)/\sqrt2$,
${\bf{p}}_y$=$({\bf{e}}_z$+${\bf{e}}_x)/\sqrt2$,
${\bf{p}}_y^b$=$(-{\bf{e}}_z$+${\bf{e}}_x)/\sqrt2$,
${\bf{p}}_z$=$({\bf{e}}_x$+${\bf{e}}_y)/\sqrt2$ and
${\bf{p}}_z^b$=$({\bf{e}}_x$$-$${\bf{e}}_y)/\sqrt2$. The E1 and E2
distributions here coincide and M1 differs from them by sign and an offset,
\begin{equation}
\begin{split}
q_{\rm E1} ({\bf{r}})=q_{\rm E2} ({\bf{r}})= \Delta q/2 &+ 2\rho_x \rho_z \sin kx\sin kz \\
&+ 2\rho_y \rho_z \cos ky\cos kz,\\
q_{\rm M1} ({\bf{r}})=\Delta q -q_{\rm E1} ({\bf{r}}).
\end{split}
\end{equation}

(III) ${\bf{E}}_f  \bot {\bf{E}}_b $ standing waves (${\bf{p}}_\xi  \bot {\bf{p}}_\xi^b $), in which  we take polarization vectors
to be ${\bf{p}}_x^ \pm$=$
{\bf{e}}_y$$\pm$${\bf{e}}_z $, ${\bf{p}}_y^ \pm $=$ {\bf{e}}_z$$
\pm$$ {\bf{e}}_x$, and ${\bf{p}}_z^ \pm$=${\bf{e}}_x $$ \pm$$
{\bf{e}}_y $.
Then the E1 and M1 distributions coincide and E2 differs from them by sign and an offset,
\begin{equation}
\begin{split}
q_{\rm E1} ({\bf{r}}) =&q_{\rm M1} ({\bf{r}}) = \Delta q/2  + \rho_x \rho_y \cos k(x + y) \\
+& \rho_y \rho_z \cos k(y + z) + \rho_z \rho_x \cos k(z + x),\\
q_{\rm E2} ({\bf{r}})=&\Delta q- q_{\rm E1} ({\bf{r}}).
\end{split}
\label{}
\end{equation}

As indicated by these examples, it is essential that the spatial  distributions of
$q_X({\bf r})$ 
may show the same spatial dependences apart from the sign and an offset $\Delta q$ for particular lattice geometries. 
However, we note that this is not a general feature
for optical lattices. For example, in the 3D lattice with
${\bf{E}}_f ||{\bf{E}}_b $ standing waves with  ${\bf{p}}_x$=$
{\bf{p}}_y$=${\bf{e}}_z$ and ${\bf{p}}_z$=${\bf{e}}_x$ employed in
our previous experiment~\cite{Aka08}, neither $q_{\rm M1}
({\bf{r}})$ nor $q_{\rm E2} ({\bf{r}})$ shows the same spatial
dependences as  $q_{\rm E1} ({\bf{r}})$, therefore the motional
effects may limit clock uncertainties in future experiments.

These three examples show prominent features by themselves.  For
a magic frequency, where the E2 (M1) interaction is
significantly larger than the  M1 (E2) interaction, case (II) [case (III)]
will be more advantageous than the others, as the less significant M1 (E2)
contribution may well be neglected.
For application to  the blue magic wavelength, which highlights the
reduction of the hyperpolarizability effects by trapping atoms near
the nodes, case (I) would be a reasonable
choice, as Eq.~(\ref{caseI}) suggests the creation of perfect nodes
regardless of the intensity balance in the orthogonal lattice beams.
Regarding the lattice light polarization,  case (I) shows linear polarization, while
cases (II) and (III) have elliptical polarizations that may give rise to the vector shift for atoms with non-zero angular momentum.

Hereafter, we focus on  case (I) and consider its application to the blue detuned magic wavelength.
The clock transition frequency   is expressed as
\begin{equation}
\begin{split}
 \nu_{\rm clock}(\omega ) = \nu _0 & - \frac{1}{2}\Delta \alpha _{\rm EM} (\omega )q_{\rm E1} ({\bf{r}}) E_0^2 \\
&  - \frac{1}{2}\Delta \alpha _{0} (\omega ) \Delta q E_0^2  + {\cal O}\left(E_0^4 \right),\\
\end{split}
\label{magic}
\end{equation}
where $\nu_0$ is the atomic transition frequency, $\omega $ is the
lattice laser frequency.
The quadratic light shift is decomposed into two parts depending on their spatial dependences by $\Delta \alpha _{\rm EM}
(\omega )\equiv \Delta \alpha _{\rm E1} (\omega )-\Delta \alpha
_{\rm M1} (\omega )-\Delta \alpha _{\rm E2} (\omega ) $ and
$\Delta \alpha _{0} (\omega )\equiv \Delta \alpha _{\rm M1}
(\omega )+\Delta \alpha _{\rm E2} (\omega ) $ using  differential
E1, M1, and E2 polarizabilities in the clock transition. Besides
the hyperpolarizability effects that are minimized by use of a
blue-detuned  lattice~\cite{Tak09}, the magic frequency
$\omega_{\rm m}$ may be given by $\Delta \alpha _{\rm EM}
(\omega_{\rm m} ) = 0$,  allowing us to  define $\omega_{\rm m}$
independent of atomic motional states. The residual M1-E2 term
$\delta \nu =- \frac{1}{2}\Delta \alpha _{0} (\omega_{\rm m} ) \Delta q
E_0^2$ provides an atomic-position-independent offset, which is solely
related to the total lattice laser intensity  $\Delta
q E_0^2=2(E_x^2+E_y^2+E_z^2)$.

To find the magic frequency, it is essential to extract
the atomic-motion-dependent term in Eq.~(\ref{magic}).
Harmonically approximating the trapping potential near the lattice
node at $x$=$y$=$z$=$\lambda/4$ and averaging over the atom
positions in the oscillator state $|\mathbf n\rangle=|n_x,n_y,n_z\rangle$, the second term in
the right-hand side of Eq.~(\ref{magic}) should be replaced wth
\begin{equation}
 2\left\langle- \frac{1}{2}\Delta \alpha _{{\rm{EM}}} q_{\rm E1}({\bf
 r})E_0^2\right\rangle_{\mathbf n} = \sum\limits_{\xi} (\Omega _\xi^{P}
 - \Omega _\xi^{S} )(n_\xi  + \frac{1}{2}),
\end{equation}
where  $\Omega _\xi^{(\ell )}$=$ k \rho_\xi E_0 \sqrt {{2  |\alpha
_{\rm EM}^{(\ell )} (\omega ) |  }/\mathcal{M}}$ is the
vibrational frequency of atoms in the $\xi$ direction of the
lattice potential for the $\ell=P (^3P_0)$ or $S(^1S_0)$ state.
Here $\mathcal M$ is the mass of the atom, $\alpha _{\rm
EM}^{(\ell )} (\omega )$$\equiv $$\alpha _{\rm E1}^{(\ell )}
(\omega )$$-$$\alpha _{\rm M1}^{(\ell )} (\omega )$$-$$\alpha
_{\rm E2}^{(\ell )} (\omega )$ is given by the E1, M1, and E2
polarizabilities in the $\ell$ state, and factor 2 accounts for the
kinetic energy.

Atomic-motion-dependent effects can be  identified by the clock
frequency difference $\Delta \nu (\omega , \delta {\bf{n}})$=
$\nu_{\rm clock} (\omega , \delta {\bf{n}}$+${\bf{n}})
$$-$$\nu_{\rm clock} (\omega ,{\bf{n}})$ measured for atoms
occupying  vibrational states differing by $\delta {\bf{n}}$.
The magic frequency $\omega_{\rm m}$ \footnote{This definition of the magic frequency is different from that in 
Ref.~\cite{Tak09} determined for a traveling wave. }
can be determined by  $\Delta
\nu (\omega_{\rm m} , \delta {\bf{n}})$=0,  equivalent to $\Delta
\alpha _{\rm EM}^{} (\omega _{\rm{m}} )$=$\alpha _{\rm EM}^{P}
(\omega _{\rm{m}} )$$-$$\alpha _{\rm EM}^{S} (\omega _{\rm{m}}
)$=0. 
Once the magic frequency is determined, the residual M1-E2
offset $\delta \nu$ can be evaluated in terms of the vibrational
frequencies of the lattice potential $\Omega _\xi$=$\Omega
_\xi^{S} (\omega_{\rm m })$=$\Omega _\xi^{P} (\omega_{\rm  m} )$ as
\begin{equation}
\delta \nu  =  - \frac{{\mathcal M\Delta \alpha _0 }}{{2k^2 \left|
{\alpha _{\rm EM} } \right|}}\left( {\Omega _x^2  + \Omega _y^2  +
\Omega _z^2 } \right).
\end{equation}
Therefore all the essential measurements are done by the frequency
measurements, once the magic frequency/wavelength is measured and shared.
The same strategy should apply to 1D optical lattices with red-detuned magic wavelength by setting $\rho_x\not=0$ and $\rho_y=\rho_z=0$ in  case (I).

The proposed optical lattice will be conveniently realized by a folded
lattice~\cite{Rau98,Aka08}, which maintains the relative phases of the
orthogonal standing waves to realize linear lattice polarizations.
Eq.~(\ref{Efield}) essentially  assumes that the
intensities of the counter-propagating beam pairs are balanced,
which is accomplished  by preparing lattice beams inside an
optical  cavity \cite{Aka08}. The blue-detuned magic wavelength
for the Sr clock transition  is experimentally determined to be
$\lambda_{\rm m}$=$2\pi c/\omega_{\rm m}$$\approx$$
389.9$~nm~\cite{Tak09}, where the numerical estimates 
for this
lattice give $\Delta \alpha_0/|\alpha_{\rm EM}|$$\approx$$
-1.4\cdot10^{-7}$ with $\Delta\alpha_{\rm M1}/
\Delta\alpha_{\rm E2}$$\approx$$ 8\cdot 10^{-3}$ \footnote {As for the
red-detuned lattice at $\lambda_{\rm m} \approx
813.4$~nm, $\Delta \alpha_0/|\alpha_{\rm
EM}|\approx-3.3\cdot10^{-8}$ with $\Delta\alpha_{\rm M1}/
\Delta\alpha_{\rm E2}\approx7\cdot 10^{-2}$.}.
Therefore, the offset frequency is given by
 $\delta \nu/2\pi$$\approx40 \cdot I$~mHz for a trap
frequency of $\Omega_\xi/2\pi$=75$\sqrt I$~kHz, where $I$ is the single running wave laser intensity measured in
${\rm kW/cm}^2$ assuming $\rho_\xi=1$. The uncertainty for this correction may be evaluated to be
$\approx 4I$~mHz, assuming an inhomogeneity of the lattice
intensity of 10~\%.

For $^{87}$Sr atoms with a total angular momentum of $F$=9/2, 
the tensor light shift due to the spatial
rotation of the lattice polarization with respect to the quantization
axis may occur. As the shift is proportional to the light
intensity near the nodes, the shift may be reduced to the mHz
level for the blue magic wavelength. Bosonic isotopes such as
$^{88}$Sr or other atomic elements with nuclear spin of $I$=1/2,
e.g., $^{171}$Yb or $^{199}$Hg may well be used to suppress the
 tensor light shift.

In summary, we present general formulae for the quadratic light shift taking
multipolar atom-field interactions into account and show that the 
spatial mismatch of the interactions in the clock transition can be treated as a
spatially constant offset $\delta \nu$ for specific lattice
geometries.
Numerical estimates are made for Sr atoms, and the relevant correction can be
  determined by trap frequency measurements with the mHz level. Combined with the blue magic
wavelength, the hyperpolarizability effect is minimized, and clock
uncertainty at the  $10^{-18}$ level will be within reach.

This work was partly supported by the Photon Frontier Network
Program, MEXT, Japan, and by the Russian Foundation for Basic
Research (RFBR grant No. 07-02-00279a). H. K. 
acknowledges M. Takamoto for useful comments.

\end{document}